\documentclass[a4paper,11pt]{article}
\pdfoutput=1
\usepackage{jheppub}
\usepackage[T1]{fontenc}

\usepackage{graphicx}%
\usepackage{caption}%
\usepackage{subcaption}%
\usepackage{multirow}%
\usepackage{amsmath,amssymb,amsfonts}%
\usepackage{xcolor}%
\usepackage{textcomp}%
\usepackage{booktabs}%
\usepackage{url}
\usepackage{orcidlink}

\newcommand{\btoxulnu}{\ensuremath{B \rightarrow X_u \ell \nu}}

\newcommand{\vub}{\ensuremath{V_{ub}}}

\newcommand{\qtwo}{\ensuremath{q^2}}
\newcommand{\mx}{\ensuremath{M_X}}
\newcommand{\elb}{\ensuremath{E_{\ell}^B}}

\title{\boldmath Mapping quark-level kinematics to hadrons in a new hybrid model of semileptonic $B$ meson decays}

\author[a]{Philipp Horak\,\orcidlink{0000-0001-9979-6501},}
\author[a]{Robert Kowalewski\,\orcidlink{0000-0002-7314-0990} }
\author[b]{and Tommy Martinov\,\orcidlink{0000-0001-7846-1913}}

\affiliation[a]{Department of Physics and Astronomy, University of Victoria,\\BC, Canada}
\affiliation[b]{INFN, Sezione di Trieste,\\Trieste, Italy}

\emailAdd{phorak@uvic.ca}
\emailAdd{kowalews@uvic.ca}
\emailAdd{tommy.martinov@ts.infn.it}

\abstract{
The need to map parton-level processes to color-neutral hadrons in a way that respects quark-hadron duality arises in several areas of physics, including in the semileptonic decays of $B$ mesons.  Integrated over large regions of phase space, the quark-level and hadron-level quantities are expected to be equal.  However, the breakdown of duality is manifest at low hadron-system invariant masses, where discrete resonances dominate.  In practice, this means independent simulations of decays to low-lying resonances and to higher-mass hadronic systems must be merged into a coherent model.  We present a novel method to combine these resonant and non-resonant components in simulations of inclusive $b \to u\ell\nu$ decays that uses an optimal transport algorithm. The method currently used in measurements of inclusive semileptonic $B$ decay branching fractions introduces unphysical features in kinematic spectra such as large discontinuities and negative yields. The optimal transport method solves both of these issues and can be easily implemented in experimental studies of $B \to X_u \ell \nu$ decays.
}


\begin{document}
\maketitle

\section{Introduction}\label{sec1}

Precise determination of Cabibbo-Kobayashi-Maskawa (CKM) elements $\lvert V_{ub}\rvert$ and $\lvert V_{cb}\rvert$ remains one of the outstanding challenges in flavor physics. These parameters are central to testing CKM unitarity and the associated CP-violating mechanisms in the Standard Model (SM). Semileptonic $B$ decays provide the most accessible way to measure these parameters. These determinations can be performed inclusively, by summing over all possible final-state hadrons in $B \rightarrow X_{c/u} \ell \nu_\ell$ decays, or exclusively, by focusing on individual channels such as $B \rightarrow D^* \ell \nu_\ell$ for $|V_{cb}|$ or $B \rightarrow \pi \ell \nu_\ell$ for $|V_{ub}|$. Inclusive and exclusive measurements have exhibited a persistent $1$--$3\sigma$ tension for over two decades, with the latest HFLAV report placing the current discrepancy at the $3\sigma$ level~\cite{HFLAV}.

Inclusive determinations of $|V_{ub}|$ rely on the Heavy Quark Expansion (HQE)~\cite{Mannel1994, Chay:1990da}, which provides reliable predictions for the fully differential decay rate integrated over large regions of phase space. Quark-hadron duality is invoked to allow these quark-level calculations to be applied to experimentally observable quantities.  Final state quarks produced by the HQE simulation must undergo fragmentation and hadronization to model observable final states.

The inclusive decay rate for $B \rightarrow X_u\ell \nu$ transitions is modeled as a smooth spectrum.
However, the physical final states include discrete low-mass resonances such as $\pi$ and $\rho$, which have been separately measured~\cite{PhysRevD.110.030001}.
The fraction of the inclusive spectrum that needs to be replaced by these exclusive decays is determined by the inclusive and exclusive semileptonic branching fractions.  In the absence of a complete theory of hadronization, it is not clear which portions of the HQE phase space should be depleted of inclusive events in the process.
The \textit{hybrid model}, first proposed in Ref.~\cite{PhysRevD.41.1496}, addresses this by combining exclusive predictions for known resonances with inclusive HQE predictions.
In the conventional implementation of the hybrid model, resonances are inserted with their measured branching fractions and form factors, while the remaining inclusive component is reweighted bin-by-bin in a way that preserves the total rate.

While this procedure has been adopted in several inclusive 
$B \to X_u \ell \nu$ analyses~\cite{Belle:2021eni, 59ws-zxbt}, it introduces unphysical discontinuities in the spectra, most noticeably in $M_X$. In this work, we present an alternative approach that instead redistributes events by minimizing the aggregate distance in kinematic space that each generated decay must be moved.  This produces smooth spectra while also maintaining better consistency with the inclusive HQE predictions.

The remainder of this paper is organized as follows: we review the theoretical framework in Section~\ref{sec:theory}, describe the simulation samples in Section~\ref{sec:samples}, introduce the conventional hybrid method and its limitations in Section~\ref{sec:conventional}, present our optimal transport approach in Section~\ref{sec:emd}, show results in Section~\ref{sec:results}, and conclude in Section~\ref{sec:conclusion}.

\section{Theoretical framework}

\label{sec:theory}
The theoretical description of inclusive semileptonic $B$ decays is based on the HQE, which relies on a local operator product expansion (OPE) in which the decay rate is expressed as a double expansion in powers of the strong coupling constant $\alpha_S$ and inverse powers of the $b$-quark mass. This quark-level calculation is expected to give reliable predictions for inclusive $B\to X_u\ell\nu$ decays when integrated over the full phase space, which can be parameterized by the squared mass of the lepton-neutrino pair $q^2$, the energy of the lepton in the $B$ rest frame $E^B_\ell$, and the invariant mass $M_X$ of the hadronic system $X_u$.  However, due to the $\sim 50$-times larger decay rate for $B\to X_c\ell\nu$ decays, determining the decay rate for $B\to X_u\ell\nu$ over a large portion of the phase space is not possible experimentally\footnote{While experimental analyses may quote partial rates over large regions, e.g., for $E^B_\ell >1 \,$GeV, the experimental sensitivity in these measurements is still predominantly in the region where $B\to X_c\ell\nu$ transitions are highly suppressed by kinematics.}.  The theoretical description of partial rates in restricted regions of phase space requires the introduction of a set of non-perturbative \textit{shape functions}. The shape functions
parametrize, at leading order, the Fermi motion of the $b$ quark inside the meson.

For the purposes of modeling the fully differential inclusive spectrum, two theoretical calculations are used to describe $B \to X_u \ell \nu$ decays. The De Fazio-Neubert (DFN) approach~\cite{DeFazio:1999ptt} includes leading-power contributions with fixed-order $\mathcal{O}(\alpha_s)$ corrections and models nonperturbative effects through a leading-order shape function. The Bosch, Lange, Neubert, Paz (BLNP) framework~\cite{Lange:2005yw} employs a factorization formula with renormalization-group-improved perturbation theory to resum large logarithms arising from scale hierarchies, and yields a treatment that smoothly connects the shape-function region to the domain where the local OPE is valid. Other theoretical approaches exist for describing $B \to X_u \ell \nu$ decays, 
including GGOU~\cite{Gambino:2007rp} and DGE~\cite{Andersen:2005mj}. As of now, these models are used to 
extract partial branching fractions integrated over restricted 
regions of phase space, rather than as fully differential event generators and are therefore not explored in this paper.

The HQE predicts both total rates and spectral moments of kinematic variables. The moments are defined as
\begin{equation}
\langle x^m \rangle = \frac{1}{\Gamma} \int x^m \frac{d\Gamma}{dx} dx,
\label{eq:spectral_moment}
\end{equation}
where $x$ is a kinematic variable and $m$ is the moment order. The first moment ($m=1$) corresponds to the mean value of the distribution, while higher moments characterize the shape and width of the distribution. These moments, such as $\langle (M_X^2)^m\rangle$, $\langle (q^2)^m\rangle$, and $\langle (E^B_\ell)^m\rangle$, can be calculated within the HQE as functions of the non-perturbative parameters. Any simulation used to model $B\to X_u\ell\nu$ decays should preserve these moments to maintain consistency with the HQE predictions.

The kinematic space of the $X_u\ell\nu$ final state can also be parameterized in terms of light-cone variables $P_{\mp}=E_X\pm|\vec{p}_X|$.  The relations to $q^2$ and $M_X$ are:
\[
\begin{aligned}
    q^2 &= m_B^2 + M_X^2 - m_B (P_- + P_+) ,\\
    M_X^2 &= P_- P_+ .
\end{aligned}
\]
The angle of the charged lepton in the $W$ rest frame relative to the recoiling $B$ direction, $\cos\theta_\ell$, is also studied, and depends on the polarization state of the virtual $W$ boson in the semileptonic decay.

\section{Simulation samples}
\label{sec:samples}

The DFN and BLNP models are implemented in EvtGen~\cite{Lange:2001uf}, where the final state quark pair is produced with $M_X\ge 2m_\pi$ and subsequently hadronized using PYTHIA~\cite{Sjostrand:2014zea}. The DFN implementation has been used in all existing experimental analyses of inclusive $B\to X_u\ell\nu$ decays. Two-dimensional kinematic distributions of events produced with the BLNP model exhibit an enhancement on the edges of the allowed phase-space (see Appendix~A in Ref.~\cite{Martinov:2025cbz} for more details). While this feature currently prohibits experimental analyses from relying on events simulated with BLNP, the model can nevertheless be used as a point of comparison for distributions obtained with the DFN model.

The presence of resonant low-mass hadronic states ($\pi$, $\rho$ and others) is a prominent feature of experimental studies of $B\to X_u\ell\nu$ decays. Their differential decay rates are determined by form factors. In modeling $B$ meson decays, these single-particle resonant final states are simulated as exclusive semileptonic decays using EvtGen, and their production in the hadronization of the inclusive models discussed above is disabled. These separately simulated inclusive and exclusive samples must be combined using a heuristic procedure to construct a complete {\em hybrid} model of $B\to X_u\ell\nu$ decay.

 The simulation samples used in this work are archived on Zenodo~\cite{zenodo}.

\section{Conventional hybrid model}
\label{sec:conventional}
\begin{figure*}
    \centering
    \includegraphics[width=1.\textwidth]{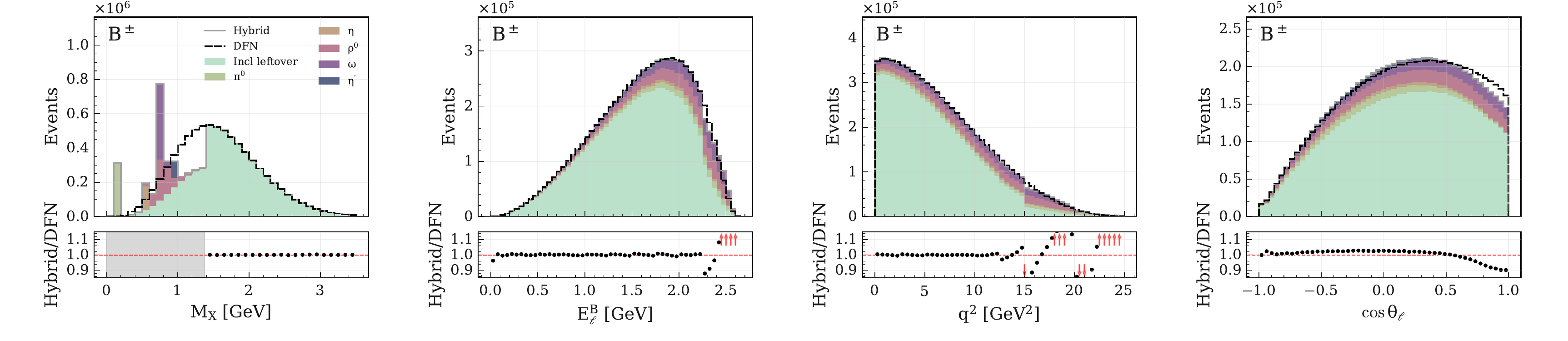}
    \includegraphics[width=1.\textwidth]{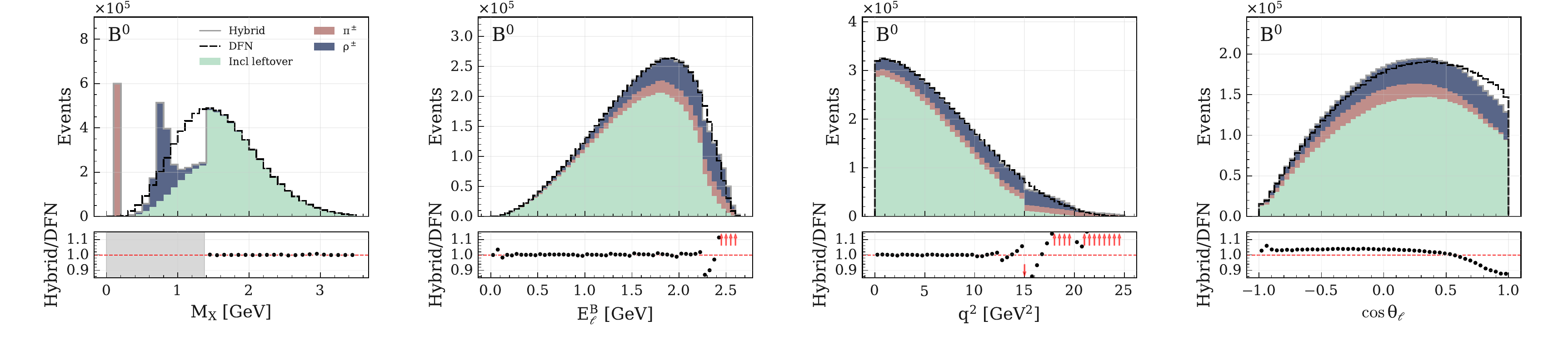}
    \caption{From left to right: distributions of \mx, \elb, \qtwo\ and $\cos\theta_\ell$ for $B^+ \to X_u^0 \ell^+ \nu$ (top) and $B^0 \to X_u^- \ell^+ \nu$ (bottom) decays after applying the hybrid weights defined in Eq.~\ref{equ:classic_hybrid_weights}. The DFN model is used to simulate inclusive decays (light green). The binning used to compute the hybrid weights is given in Ref.~\cite{59ws-zxbt}.}
    \label{fig:classic_hybrid}
\end{figure*}
The combination of resonant and non-resonant decays is typically performed via a bin-by-bin reweighting method~\cite{PhysRevD.41.1496}. The two components are summed and the non-resonant contributions are scaled down to preserve the total inclusive branching fraction. The weights applied to the non-resonant component are computed as
\begin{equation}\label{equ:classic_hybrid_weights}
    w_i = \frac{\Delta \mathcal{B}^{\textrm{inc}}_i - \Delta \mathcal{B}^{\textrm{exc}}_i}{\Delta \mathcal{B}^{\textrm{inc}}_i},
\end{equation}
where $\Delta\mathcal{B}^{\textrm{exc}}_i$ and $\Delta\mathcal{B}^{\textrm{inc}}_i$ are the initially chosen exclusive and inclusive branching fractions, respectively, in bin $i$ of the selected phase space. Recent measurements have used three-dimensional binnings in \qtwo, \elb, and \mx\ to compute these weights~\cite{Belle:2021eni, 59ws-zxbt}.

Figure~\ref{fig:classic_hybrid} shows kinematic distributions after applying this hybridization method for $B^+ \to X_u^0 \ell^+ \nu$ and $B^0 \to X_u^- \ell^+ \nu$ decays using the DFN model and the binning from Ref.~\cite{59ws-zxbt}.  The resulting distribution of $\cos\theta_\ell$, which is not used in the reweighting, is also shown.

This conventional approach has two main limitations. First, when the exclusive contribution exceeds the inclusive prediction in a given bin, the weight $w_i$ becomes negative. While a few small negative weights can be set to zero without significant impact, large or numerous negative weights -- as observed when using the BLNP model -- lead to unphysical kinematic distributions that cannot be easily corrected.

Second, the method produces visible discontinuities at bin boundaries. This is most apparent at the edge of the first \mx\ bin (at $M_X = 1.4$~GeV in Figure~\ref{fig:classic_hybrid}), which contains all measured $X_u$ resonant states. Similar discontinuities, though less pronounced, are also visible in the \qtwo\ distribution. Currently, the experimental resolution limits the impact of these discontinuities on measurements. As detector performance and reconstruction techniques improve, these artifacts become increasingly problematic.

Both issues stem from the same root cause: the bin-by-bin approach treats each region independently, without considering the global phase space. The optimal transport method presented in the following section addresses these limitations by finding a smooth, global redistribution of probability mass.

\section{Optimal transport hybrid}
\label{sec:emd}
\begin{figure*}
    \centering
    \includegraphics[width=1.\textwidth]{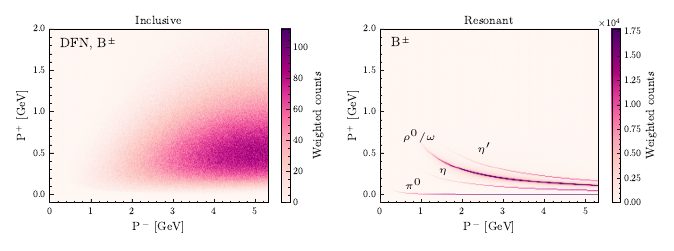}
    \caption{$P_+$ versus $P_-$ distributions for inclusive DFN (left) and resonant (right) $B^+ \to X_u^0 \ell^+ \nu$ events. The black line indicates the edge of the physical range where $M_X = 0$. The lines on the resonant plot correspond to decays to individual resonances. }
    \label{fig:source_target}
\end{figure*}

To overcome the limitations of the standard bin-by-bin reweighting, where the necessary coarse binning leads to visible discontinuities, we treat the problem as one of optimal transport.
Rather than adjusting each bin independently, we look for an optimal global mapping that redistributes probability mass from the inclusive prediction to a target distribution constructed from the known resonances, minimizing the overall \textit{cost} of moving events across the $P_-\,P_+$ phase space. In principle, this redistribution is an event-by-event (unbinned) process. However, to keep the computation manageable, we use a finely binned grid for the source and target distributions. The inclusive DFN and resonant distributions in the $P_-\,P_+$ plane for $B^+ \to X_u^0 \ell^+ \nu$ events are shown in Figure~\ref{fig:source_target}.

The difference between two distributions can be measured using the Wasserstein distance of order $p$. For two normalized discrete distributions
\begin{equation}
    \mu = \{\mu_i\}, \qquad \nu = \{\nu_j\},
\end{equation}
defined over kinematic bins, the $p$-Wasserstein distance is
\begin{equation}
    W_p(\mu, \nu) = \left( \min_{T \in \Pi(\mu,\nu)} \sum_{i,j} T_{ij}\, C_{ij}^p \right)^{1/p},
\end{equation}
where $T_{ij}$ is a \textit{transport plan} that satisfies
\begin{equation}
    \sum_j T_{ij} = \mu_i, \qquad \sum_i T_{ij} = \nu_j,
\end{equation}
and $C_{ij}$ is the distance between bins $i$ and $j$. For the case of $p=1$, $W_1$ is known as the earth-mover's distance (EMD): the minimal ``work'' needed to transform one distribution into another, analogous to the amount of soil that must be moved to reshape one pile into another. The corresponding optimal transport plan then specifies exactly how that soil should be moved.

In the \btoxulnu\ hybrid model, the transport plan gives, at each point of the inclusive $P_-\, P_+$ distribution, the fraction of the probability mass which needs to be moved elsewhere to optimally reproduce the resonant distribution.  The hybrid is constructed using the transport plan only to assign weights (the fraction of mass that would be left behind after transport) to the generated inclusive events, rather than to physically displace them.

\subsection{Implementation}\label{sec:implementation}
To implement the optimal transport mapping, we discretize the relevant phase space using a two-dimensional grid in $P_+$ and $P_-$, with a bin size small enough to preserve the relevant kinematic features. The source distribution $\mu$ is taken from the inclusive prediction, while the target distribution $\nu$ is constructed from the sum of known resonances. Since the EMD is defined for distributions with equal total mass, we introduce a \textit{sink node} to absorb source contributions that exceed the total target mass. Entries assigned to the sink node correspond to what remains of the inclusive distribution after the resonant modes have been fully populated, ensuring that the total inclusive rate is preserved.
The final hybrid model is given by summing the residual inclusive distribution with the resonant modes,
\begin{equation}\label{equ:hybrid_model}
    \mathcal{H}_i = \mu_{{\rm resid},i} + \nu_i,
\end{equation}
where $\mu_{{\rm resid},i}$ is the residual inclusive contribution in bin $i$ after subtracting the contributions assigned to the resonant target distribution, and $\nu_i$ is the resonant target distribution. 
In practice, a hybrid weight $w_i = \mu_{{\rm resid},i}/\mu_i$ is applied to the inclusive prediction: $\mathcal{H}_i = w_i\,\mu_i + \nu_i$. These weights can be used to produce distributions of any quantity of interest in simulated \btoxulnu\ decays. 

The cost matrix $C_{ij}$ quantifies the distance between source and target bins in phase space. Different choices of distance metric are possible depending on the physics considerations. In this work, we use the Euclidean distance between bin centers,
\begin{equation}
    C_{ij} = \sqrt{ (P_{+,i}^{\mu} - P_{+,j}^{\nu})^2 + (P_{-,i}^{\mu} - P_{-,j}^{\nu})^2 },
\end{equation}
where $P_{+,i}^{\mu}$ and $P_{-,i}^{\mu}$ are the coordinates of the $i$-th source bin, and $P_{+,j}^{\nu}$ and $P_{-,j}^{\nu}$ are the coordinates of the $j$-th target bin. The cost to move an event from any source bin to the sink node is set to zero. The choice of Euclidean distance with Wasserstein order $p=1$ was tested against alternative cost functions, such as the squared Euclidean distance ($p=2$), and was adopted due to its superior moment conservation (see Section~\ref{sec:results}). Alternative cost metrics could include mode-dependent cost factors to preferentially preserve certain exclusive channels, or local smoothing based on neighboring bin populations to favor gradual transitions in the transport plan. 

The optimal transport plan $T_{ij}$ is obtained numerically using the Python Optimal Transport (POT) library~\cite{JMLR:v22:20-451}. Our implementation is made available on GitHub~\cite{github}. As a cross-check, we have verified that the Google OR-Tools MinCostFlow solver~\cite{ortools} produces results consistent with the POT implementation. For the exact solution, we use the network simplex algorithm~\cite{Orlin1997}. Alternatively, an entropic regularization term can be added to the optimization problem, transforming it into
\begin{equation}\label{equ:sinkhorn}
    W_{1,\lambda}(\mu, \nu) = \min_{T \in \Pi(\mu,\nu)} \left[ \sum_{i,j} T_{ij}\, C_{ij} + \lambda H(T) \right],
\end{equation}
where $H(T) = \sum_{i,j} T_{ij} \log T_{ij}$ is the entropy of the transport plan and $\lambda > 0$ controls the strength of regularization. This regularized problem can be solved more efficiently using the Sinkhorn algorithm~\cite{sinkhorn}. The entropic term encourages smoother, more diffuse transport plans at the cost of increased transport distance.

\section{Results}
\label{sec:results}
We apply the EMD minimization to several configurations, in both $B^+ \rightarrow X_u^0 \ell^+ \bar{\nu}_\ell$ and $B^0 \rightarrow X_u^- \ell^+ \bar{\nu}_\ell$ decays, using either the DFN or BLNP parameterization for the inclusive prediction. We present here results obtained with the DFN model; additional configurations are discussed in the Appendix.

Table~\ref{tab:excl} summarizes the branching fractions for exclusive final states taken from the Heavy Flavor Averaging Group review~\cite{HFLAV}. The inclusive rate is taken separately for neutral and charged $B$ decays as quoted in the latest Review of Particle Physics~\cite{PhysRevD.110.030001}. The inclusive branching fractions are measured for $\elb > 1$~GeV and scaled to the full phase space. Parameters of the DFN model are taken from Ref.~\cite{PhysRevD.73.073008} in the Kagan-Neubert renormalization scheme~\cite{Kagan1999}. The BLNP parameters are obtained in the shape function scheme~\cite{NEUBERT200513} as described in Ref.~\cite{HFLAV} and then converted to the shape function parameters as given in Ref.~\cite{Lange:2005yw}.
\begin{table}
    \centering
    \captionsetup{width=\linewidth}
    \caption{Inputs to the hybrid construction. (a) Exclusive and inclusive 
    branching fractions. (b) Parameters of the inclusive models. For details 
    about the DFN and BLNP parameters, see Refs.~\cite{DeFazio:1999ptt} 
    and~\cite{Lange:2005yw}, respectively.}
    \label{tab:hybrid_inputs}
    \begin{subtable}[t]{0.45\linewidth}
        \centering
        \caption{}
        \label{tab:excl}
        \begin{tabular}{ccc}
            \toprule\toprule
            Decay mode & \multicolumn{2}{c}{$\mathcal{B}~[10^{-4}]$}\\[2pt]
                       & $B^+$ & $B^0$ \\
            \midrule
            $B \rightarrow \pi\,\ell^+\nu$      & $0.78$ & $1.50$ \\
            $B \rightarrow \rho\,\ell^+\nu$     & $1.58$ & $2.94$ \\
            $B \rightarrow \omega\,\ell^+\nu$   & $1.19$ &        \\
            $B \rightarrow \eta\,\ell^+\nu$     & $0.35$ &        \\
            $B \rightarrow \eta'\,\ell^+\nu$    & $0.24$ &        \\
            \midrule
            $B \rightarrow X_u \ell^+\nu$       & $19.2$ & $17.6$ \\
            \bottomrule\bottomrule
        \end{tabular}
    \end{subtable}
    \hfill
    \begin{subtable}[t]{0.45\linewidth}
        \centering
        \caption{}
        \label{tab:models}
        \begin{tabular}{lc|cc}
            \toprule\toprule
            Parameter & DFN & Parameter & BLNP \\
            \midrule
            $m_b$     & $4.66$ & $b$         & $4.23$ \\
            $a$       & $1.33$ & $\Lambda$   & $0.75$ \\
                      &        & $\mu_h$     & $0.44$ \\
                      &        & $\mu_i$     & $1.5$  \\
                      &        & $\bar{\mu}$ & $1.5$  \\
            \bottomrule\bottomrule
        \end{tabular}
    \end{subtable}

\end{table}

\begin{figure}
    \centering
    \includegraphics[scale=0.5]{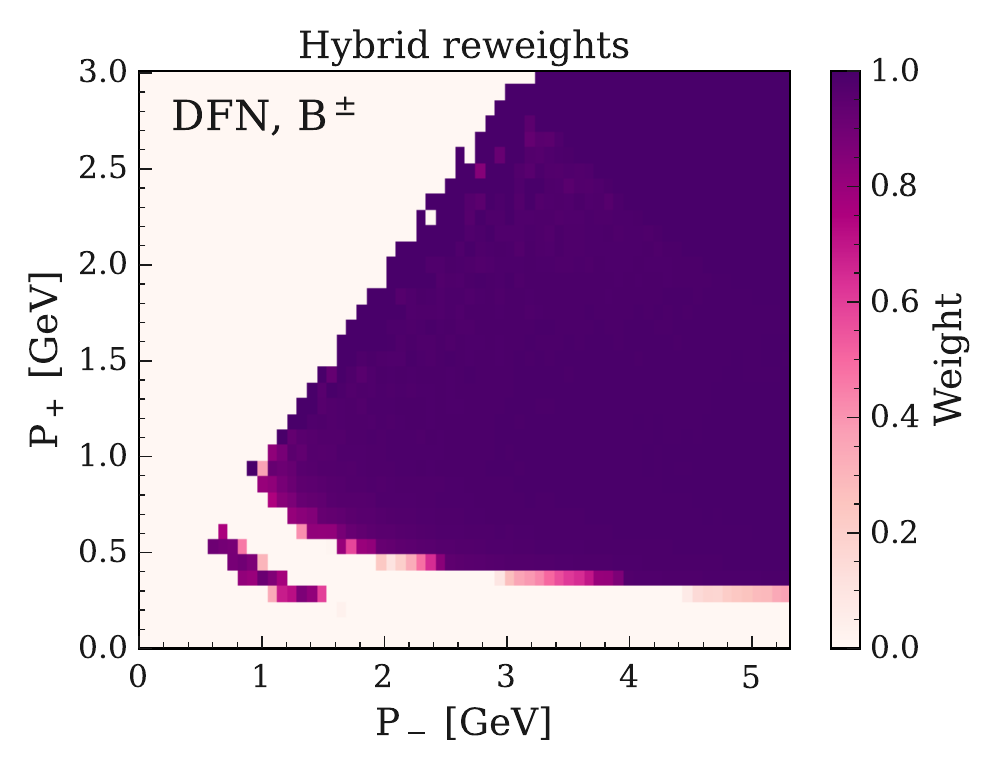}
    \caption{Optimal transport hybrid weights in the $P_- P_+$ plane for $B^+ \to X_u^0 \ell^+ \nu$ decays simulated with the DFN model. Bins with weights close to unity (dark purple) are largely unaffected by the hybridization, while the light-colored band at low $P_+$, which corresponds to the resonant region, is mostly depleted.}
    \label{fig:reweights}
\end{figure}

\begin{figure*}
    \centering
    \includegraphics[width=1.\textwidth]{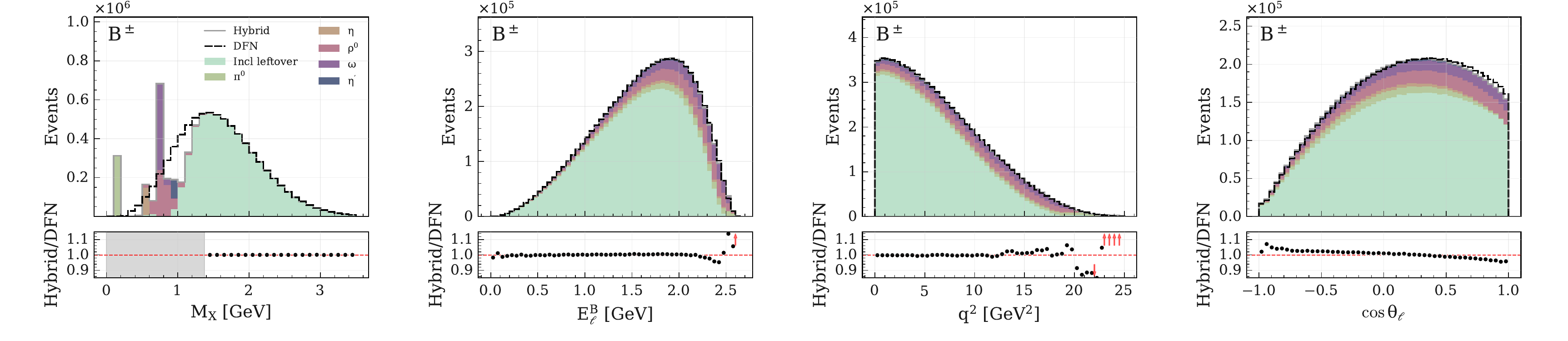}
    \includegraphics[width=1.\textwidth]{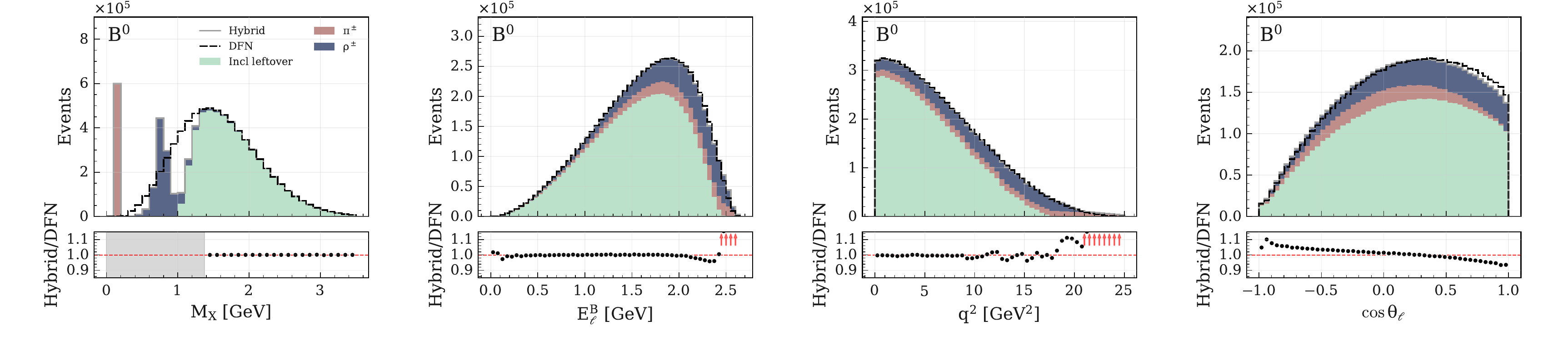}
    \caption{As Fig.~\ref{fig:classic_hybrid}, but using the optimal transport 
    hybrid weights with bins of width 0.08~GeV in $P_+$ and $P_-$.}
    \label{fig:emd_hybrid}
\end{figure*}

Figure~\ref{fig:source_target} shows the source and target distributions in the $P_-\,P_+$ plane constructed from these inputs. For visualization purposes, a finer binning is used in this figure than in the actual minimization procedure, allowing individual resonances to be more clearly resolved.
% The EMD minimization yields the transport plan $T_{ij}$, from which a residual 
% inclusive distribution $\mu_{\rm resid}$ is obtained after the resonant modes have 
% been fully populated. The hybrid weight for bin $i$ is then
% \begin{equation}\label{equ:emd_hybrid_weights}
%     w_i = \frac{\mu_{{\rm resid},i}}{\mu_{i}}.
% \end{equation}
% These weights are shown in the $P_-\,P_+$ plane for the $B^+ \to X_u^0 \ell^+ \nu$ channel and the DFN model in Figure~\ref{fig:reweights}.
The hybrid weights $w_i$ defined in Sec.~\ref{sec:implementation} are evaluated from the resulting transport plan $T_{ij}$ and shown in the $P_-\,P_+$ plane in Figure~\ref{fig:reweights}.
The resulting weight tables are provided with the code~\cite{github}. The kinematic distributions of the final hybrid model, obtained by summing the residual source and the target distributions, are shown in Figure~\ref{fig:emd_hybrid}. \\

\noindent Comparison with the bin-by-bin hybrid reveals notable improvements: the $M_X$ spectrum no longer exhibits the unphysical discontinuity that appears in the bin-by-bin approach due to the use of a wide lower bin to cover the resonant region and avoid negative weights, and the smaller discontinuity at $q^2 \simeq 14$~GeV$^2$ present in the bin-by-bin hybrid is also eliminated.
As a cross-check, we compare the lepton helicity angle distribution, $\cos\theta_\ell$, which is not used as an input to either hybridization procedure. The optimal transport hybrid does not introduce extra distortions in this variable.\\

The optimal transport approach is particularly advantageous for the BLNP parameterization, where the bin-by-bin hybrid method suffers from a large number of negative weights. The optimal transport hybrid successfully constructs a physically meaningful distribution in this case. For the BLNP hybrid distributions see Appendix~\ref{app:blnp}.\\

\begin{figure*}
    \centering
    \begin{subfigure}{\linewidth}
        \centering
        \includegraphics[width=\linewidth]{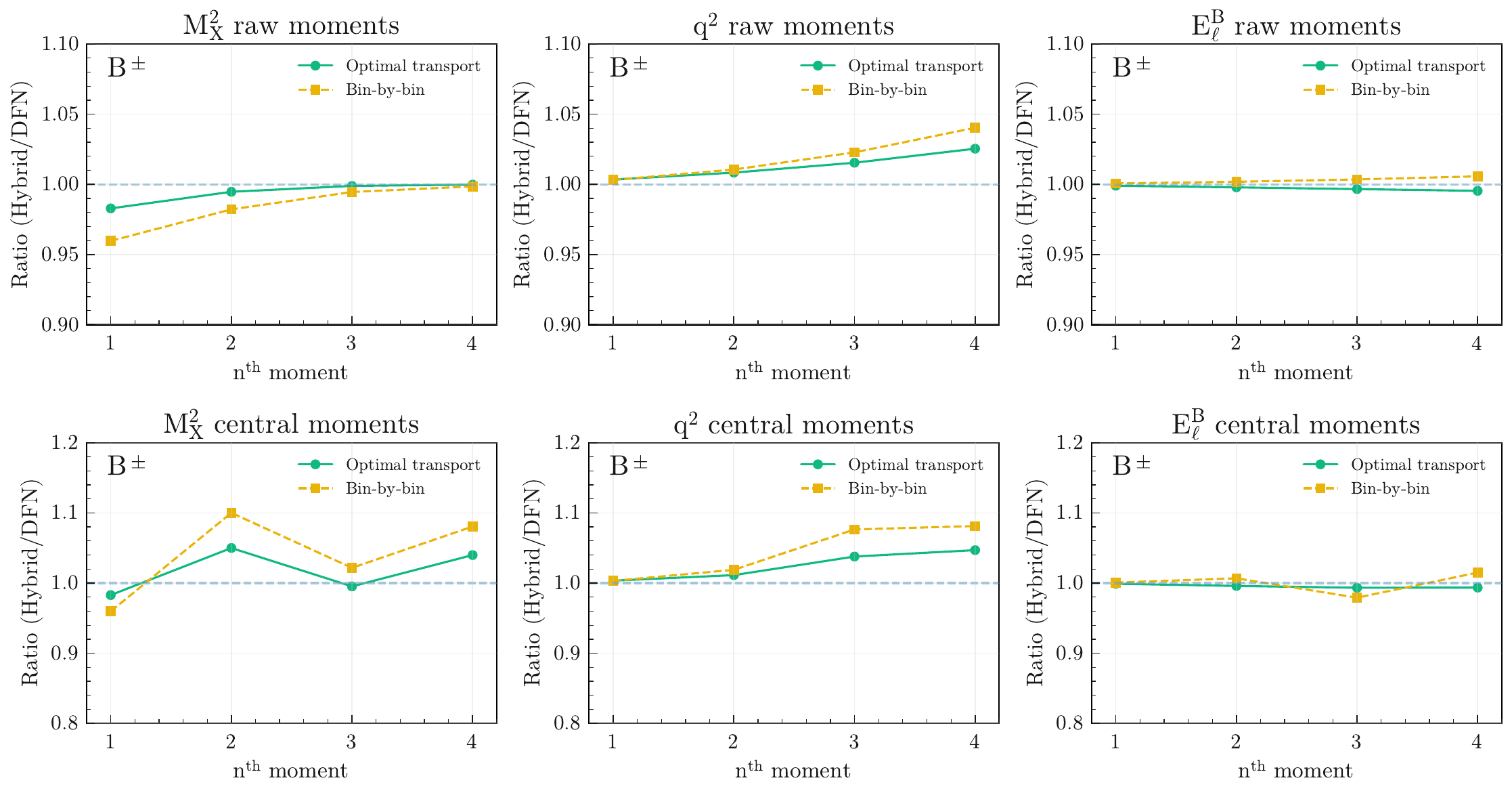}
        \caption{$B^+ \to X_u^0 \ell^+ \nu_\ell$}
        \label{fig:moments-charged}
    \end{subfigure}

    \vspace{1em}

    \begin{subfigure}{\linewidth}
        \centering
        \includegraphics[width=\linewidth]{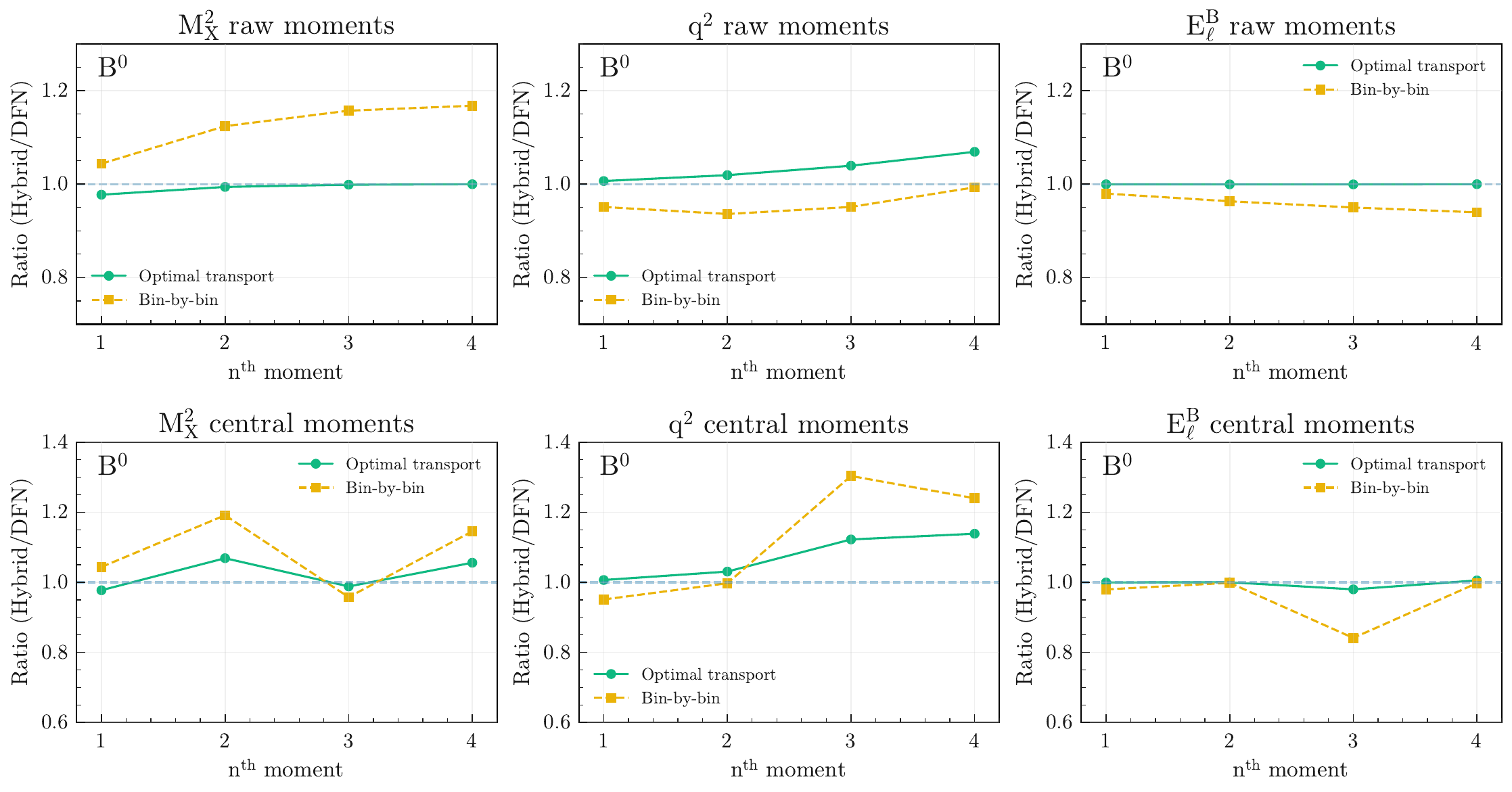}
        \caption{$B^0 \rightarrow X_u^- \ell^+ \nu_\ell$}
        \label{fig:moments-neutral}
    \end{subfigure}

    \caption{Moment conservation comparison between the optimal transport hybrid (solid green) and bin-by-bin hybrid (dashed yellow) for the (a) charged and (b) neutral modes. Top and bottom rows show raw and central moments of $M_X^2$, $q^2$, and $E_\ell^B$, plotted as the ratio to the DFN inclusive prediction. Full numerical values are given in Tables~\ref{tab:charged-moments} and~\ref{tab:neutral-moments}.}
    \label{fig:moments-both}
\end{figure*}

To quantify the degree to which the optimal transport hybrid preserves the underlying inclusive prediction, we compute the moments $\langle M_X^{2n}\rangle$, $\langle q^{2n}\rangle$, and $\langle E_\ell^n\rangle$ for both hybrid models. Figures~\ref{fig:moments-charged} and \ref{fig:moments-neutral} show the ratio of these hybrid moments to their DFN values for the charged and neutral $B$ modes, respectively. Averaged over $B$ modes and the first four moments, the optimal transport hybrid conserves the raw moments roughly four times better than the conventional approach (mean relative error of 1.0\% vs.\ 4.1\%). Breaking it down by variable, the optimal transport hybrid improves $\langle M_X^2\rangle$ from 7.0\% to 0.7\%, $\langle q^2\rangle$ from 3.1\% to 2.3\%, and $\langle E_\ell^B\rangle$ from 2.2\% to 0.2\%.\\

Finally, we note that the rate of $s\bar{s}$ pair production during hadronization can be controlled in PYTHIA through the \texttt{StringFlav:probStoUD} parameter. In our inclusive sample, the suppression of $s\bar{s}$ pairs relative to $u\bar{u}/d\bar{d}$ pairs is set to 0.217, following the Monash tune~\cite{Skands:2014pea}. The optimal transport hybrid better conserves the fraction of events in which a kaon is found. For $B^\pm$ and $\bar{B}^0$ decays simulated with the DFN model, kaons are produced in about 10.5\% of decays. After hybridization, this fraction decreases to approximately 8.5\% in the bin-by-bin hybrid but remains approximately 10.0\% in the optimal transport hybrid.\\

To explore the dependence of results on the details of the algorithm, we introduced entropic regularization into the OT problem (Eq.~\ref{equ:sinkhorn}) and solved it using the Sinkhorn algorithm, where $\lambda$ controls the smoothness of the resulting transport plan. For $\lambda \gtrsim 1$, the \mx\ distribution retains a non-resonant contribution below 1~GeV, as expected from the inclusive kinematic distributions and similarly to the bin-by-bin hybrid model. This contrasts with the optimal transport hybrid, in which resonant decays dominate this region. The kinematic distributions reweighted using entropic regularization are shown in Figure~\ref{fig:sinkhorn_hybrid} and the resulting moments are shown in Appendix~\ref{app:sinkhorn}. We note that the true composition of the \btoxulnu\ spectrum is not experimentally known, as the finite detector resolution does not allow a clean separation of resonant and non-resonant decays. The entropic regularization term offers some flexibility to tune the inclusive spectrum composition based on new results or modeling assumptions. However, the regularized spectra show weaker moment conservation and poorer agreement with the inclusive spectra compared to the optimal transport hybrid.
\begin{figure*}
    \centering
    \includegraphics[width=1.\textwidth]{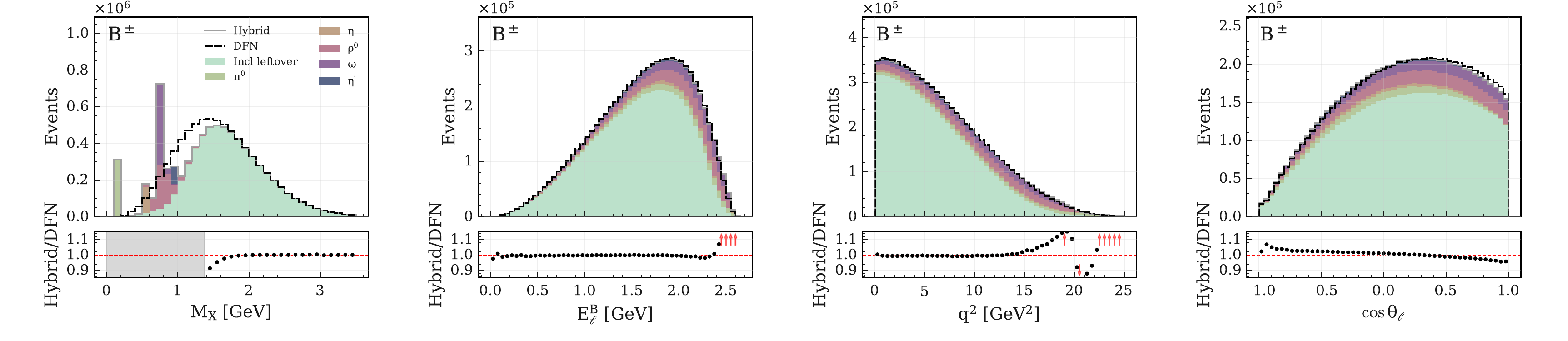}
    \includegraphics[width=1.\textwidth]{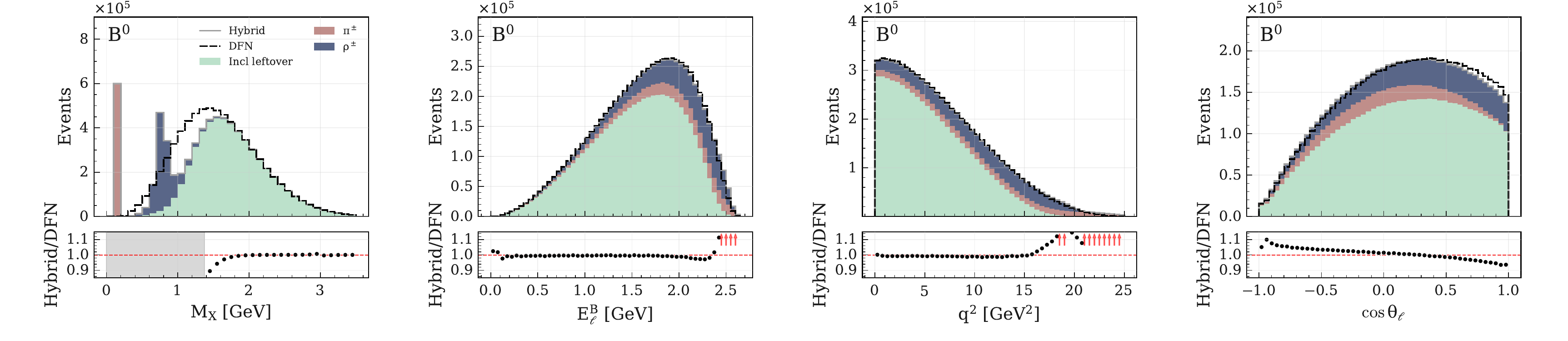}
\caption{As Fig.~\ref{fig:classic_hybrid}, but using hybrid weights obtained 
with entropic regularization ($\lambda = 1$) and bins of width 0.08~GeV 
in $P_+$ and $P_-$.}
    \label{fig:sinkhorn_hybrid}
\end{figure*}

\section{Summary}
\label{sec:conclusion}

We have presented an alternative approach to the construction of hybrid models for inclusive semileptonic $B$ decays based on optimal transport theory. Unlike the conventional bin-by-bin reweighting method, which produces unphysical discontinuities, the optimal transport hybrid finds a global redistribution of probability mass that minimizes the aggregate displacement in kinematic space.

The key advantages of this approach are threefold. First, it eliminates the unphysical discontinuities that appear at bin boundaries in the conventional method, particularly in the $M_X$ and $q^2$ distributions. Second, it naturally handles cases where the exclusive contribution locally exceeds the inclusive prediction -- a scenario that leads to negative weights in the bin-by-bin approach and prohibits the use of the current BLNP model. Third, it better preserves the spectral moments, reducing the mean relative error from 4.1\% to 1.0\% across kinematic variables and moment orders.  Incorporating new or updated measurements of exclusive semileptonic decay modes, or of new models for inclusive semileptonic decays, is straightforward.

These improvements are particularly relevant given recent experimental developments. Belle~II has recently published the first inclusive $|V_{ub}|$ measurement based on 365 fb$^{-1}$ of data~\cite{59ws-zxbt}, achieving competitive precision with a dataset representing less than 1\% of the experiment's ultimate goal of 50 ab$^{-1}$. With improving detector resolution and reconstruction capabilities at Belle~II and future facilities, the artifacts introduced by bin-by-bin reweighting are becoming increasingly problematic for precision measurements of $|V_{ub}|$. The optimal transport hybrid provides a theoretically motivated and practically superior alternative that maintains better consistency with HQE predictions while producing physically realistic kinematic distributions suitable for these next-generation analyses.

Beyond $B \to X_u \ell \nu$ decays, this framework could be naturally extended to other processes requiring hybrid modeling, such as $B \to X_s \ell^+ \ell^-$, $B \to X_s \nu\bar{\nu}$ and $B \to X_s \gamma$ decays, where similar considerations of resonant and non-resonant contributions arise. The method is general and can be applied wherever smooth interpolation between two distinct predicted distributions is needed.\\

\acknowledgments The authors are grateful to Tom Blake and Markus Prim for helpful comments on the manuscript. This work was supported in part by NSERC, Canada.\\

\appendix

\section{BLNP hybrid distributions}\label{app:blnp}
\begin{figure*}[h!]
    \centering
    \includegraphics[width=1.\textwidth]{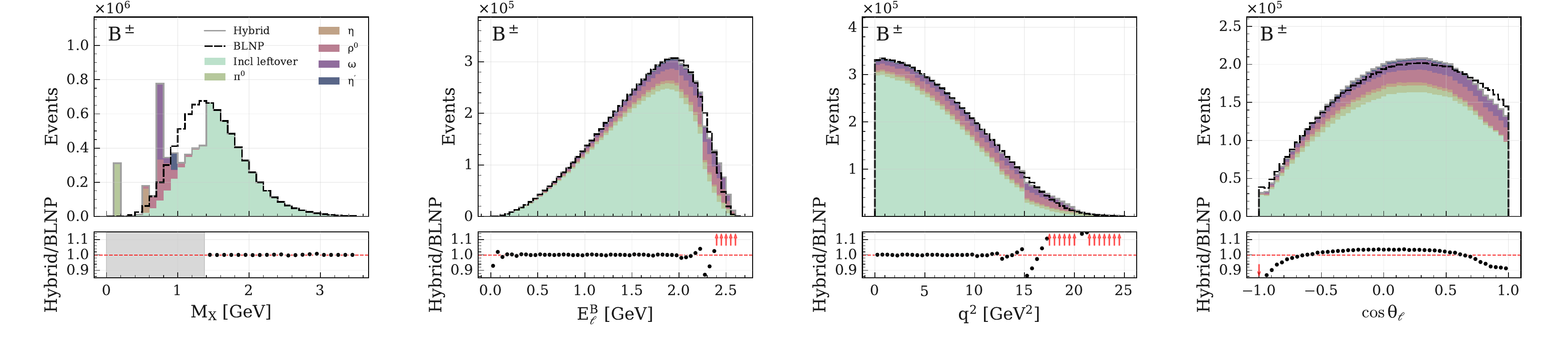}
    \includegraphics[width=1.\textwidth]{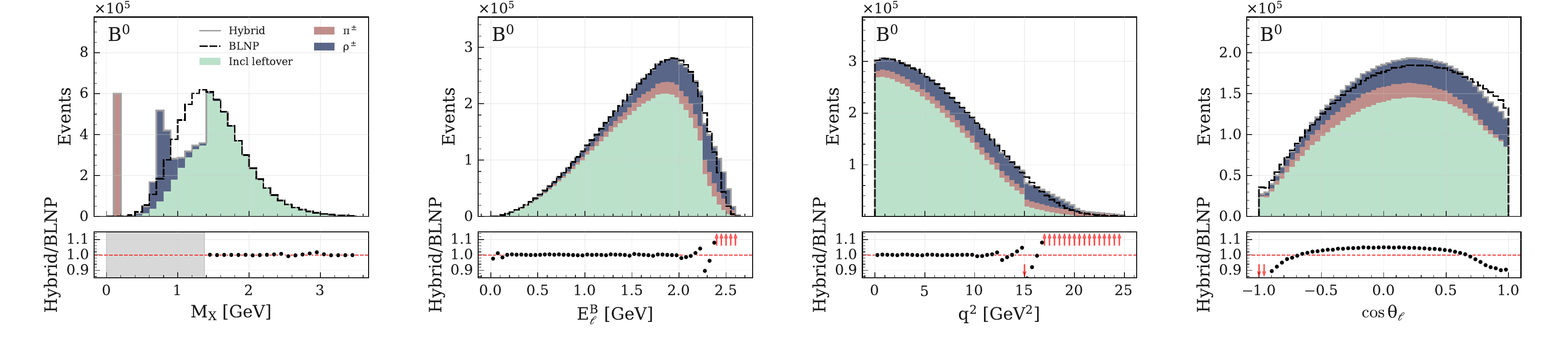}
    \caption*{(a) Conventional hybrid}
    \includegraphics[width=1.\textwidth]{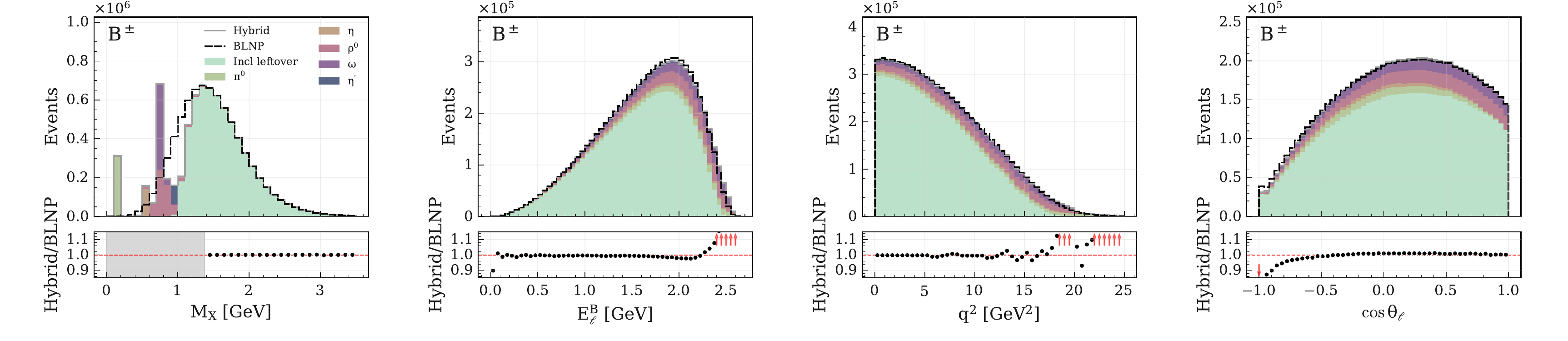}
    \includegraphics[width=1.\textwidth]{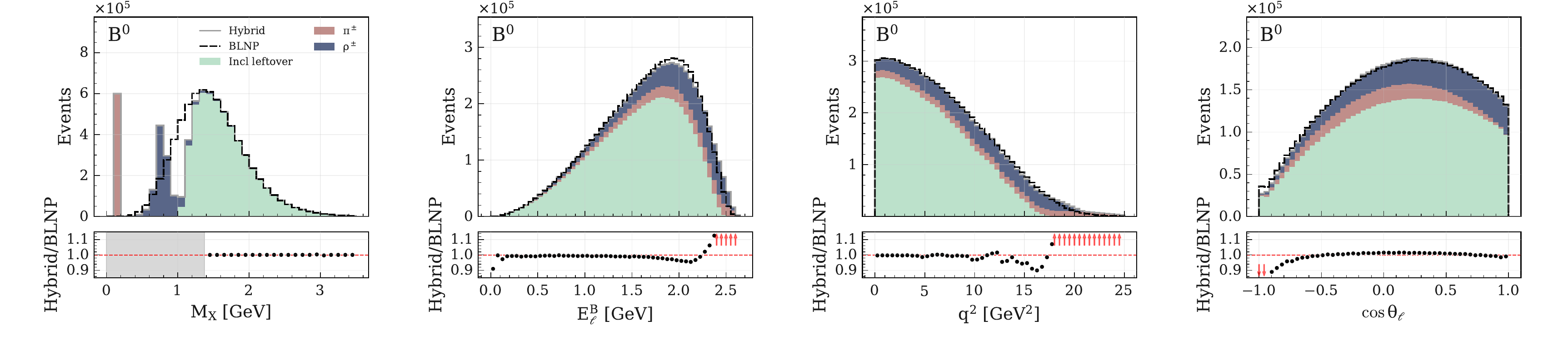}
    \caption*{(b) Optimal transport hybrid}
    \caption{As Figs.~\ref{fig:classic_hybrid} and~\ref{fig:emd_hybrid}, but using the BLNP model.}
    \label{fig:blnp_hybrid}
\end{figure*}

The kinematic distributions obtained with the conventional hybrid method and the optimal 
transport hybrid when using the BLNP model for the inclusive prediction are shown in 
Figures~\ref{fig:blnp_hybrid}a and~\ref{fig:blnp_hybrid}b, respectively.

\section{Entropic regularization moments}\label{app:sinkhorn}
\begin{figure*}[h!]
    \centering
    \begin{subfigure}{\linewidth}
        \centering
        \includegraphics[width=\linewidth]{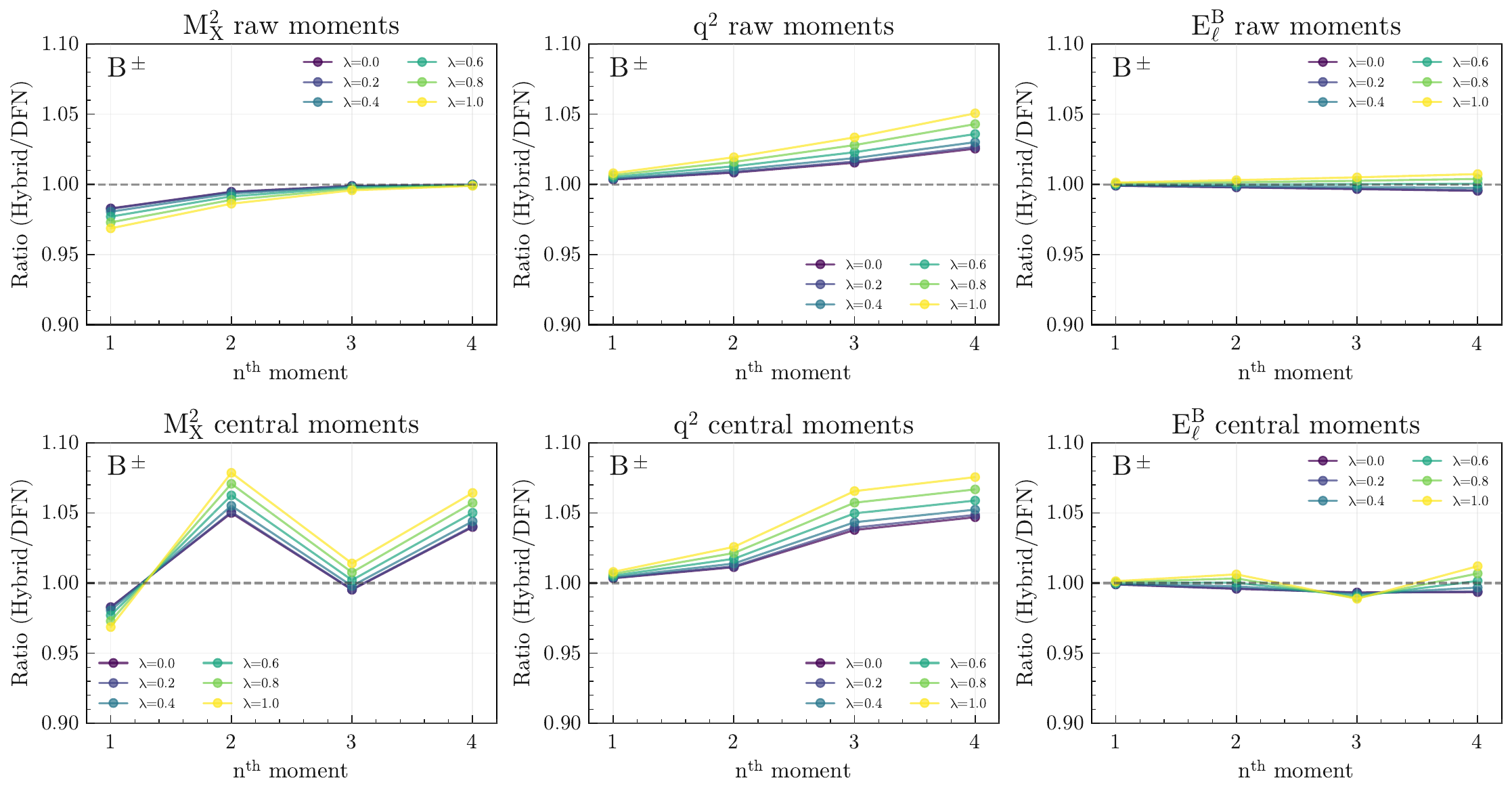}
        \caption{$B^+ \to X_u^0 \ell^+ \nu$}
        \label{fig:charged_sinkhorn_moments}
    \end{subfigure}
    \vspace{1em}
    \begin{subfigure}{\linewidth}
        \centering
        \includegraphics[width=\linewidth]{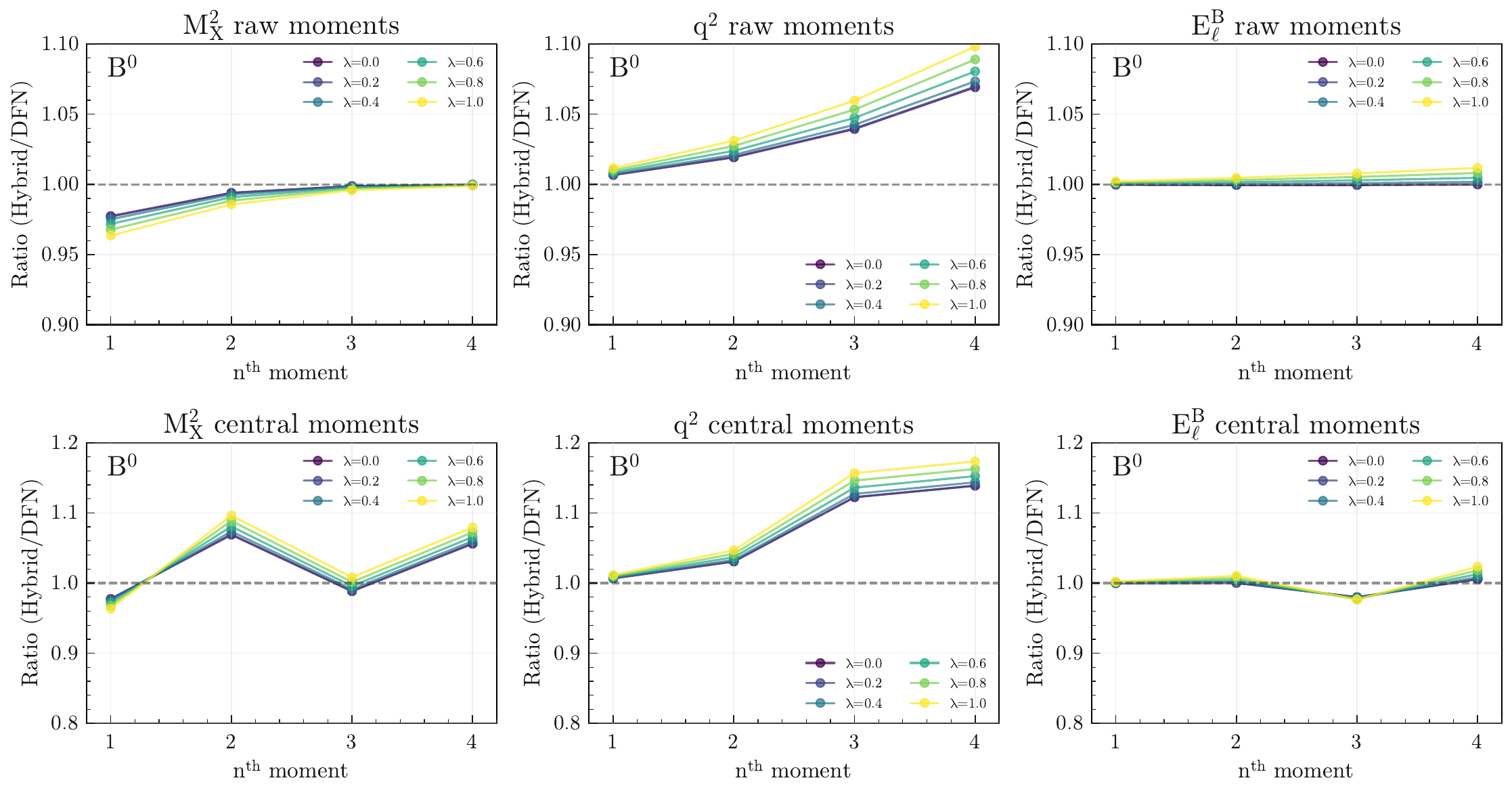}
        \caption{$B^0 \to X_u^- \ell^+ \bar{\nu}_\ell$}
        \label{fig:neutral_sinkhorn_moments}
    \end{subfigure}
    \caption{Comparison of moment conservation for the (a) charged and (b) neutral $B$ modes 
    between the optimal transport hybrid with different values of $\lambda$. Top and bottom 
    rows show raw and central moments of $M_X^2$, $q^2$, and $E_\ell^B$, plotted as the 
    ratio to the DFN inclusive prediction.}
    \label{fig:sinkhorn_moments}
\end{figure*}

The impact of the regularization strength on moment conservation is shown in Figs.~\ref{fig:charged_sinkhorn_moments} and \ref{fig:neutral_sinkhorn_moments}. As $\lambda$ increases, the conservation of moments deteriorates for all kinematic variables considered. For the distributions (Fig.~\ref{fig:sinkhorn_hybrid}), the agreement between the inclusive DFN prediction and the hybrid is visibly worse in \mx, \elb\ and \qtwo\ compared to the optimal transport hybrid. Increasing $\lambda$ enhances the smoothing of the transport plan but does not improve agreement with the inclusive prediction.

\section{Numerical moment values}
Tables~\ref{tab:charged-moments} and~\ref{tab:neutral-moments} provide the numerical values of the raw and central moments shown graphically in Figures~\ref{fig:moments-charged} and~\ref{fig:moments-neutral}. These values quantify the degree to which each hybrid method preserves the kinematic distributions predicted by the DFN model.

\begin{table*}[!h]
\centering
\caption{Moment comparison for $B^+ \rightarrow X_u^0 \ell^+ \nu_\ell$. DFN vs.~Bin-by-bin vs.~ Optimal transport hybrids.}
\resizebox{\textwidth}{!}{
\begin{tabular}{lcccccc}
\toprule \toprule
& \multicolumn{3}{c}{Raw moments} & \multicolumn{3}{c}{Central moments} \\
\cmidrule(lr){2-4} \cmidrule(lr){5-7}
Observable & DFN & Bin-by-bin & Optimal transport & DFN & Bin-by-bin & Optimal transport \\
\midrule
$\langle (M_X^2)^{1} \rangle$ & 2.83 & 2.71 & 2.78 & 2.83 & 2.71 & 2.78 \\
$\langle (M_X^2)^{2} \rangle$ & 12.15 & 11.93 & 12.08 & 4.15 & 4.57 & 4.36 \\
$\langle (M_X^2)^{3} \rangle$ & 70.81 & 70.42 & 70.73 & 12.97 & 13.25 & 12.91 \\
$\langle (M_X^2)^{4} \rangle$ & 523.84 & 523.06 & 523.75 & 113.92 & 123.10 & 118.47 \\
\midrule
$\langle (q^2)^{1} \rangle$ & 6.66 & 6.68 & 6.68 & 6.66 & 6.68 & 6.68 \\
$\langle (q^2)^{2} \rangle$ & 67.11 & 67.82 & 67.68 & 22.78 & 23.21 & 23.04 \\
$\langle (q^2)^{3} \rangle$ & 829.96 & 848.89 & 842.74 & 79.74 & 85.83 & 82.75 \\
$\langle (q^2)^{4} \rangle$ & 11655.42 & 12125.37 & 11951.64 & 1506.69 & 1629.09 & 1577.37 \\
\midrule
$\langle (E_\ell^B)^{1} \rangle$ & 1.62 & 1.62 & 1.62 & 1.62 & 1.62 & 1.62 \\
$\langle (E_\ell^B)^{2} \rangle$ & 2.87 & 2.88 & 2.87 & 0.25 & 0.25 & 0.25 \\
$\langle (E_\ell^B)^{3} \rangle$ & 5.41 & 5.43 & 5.39 & -0.06 & -0.05 & -0.05 \\
$\langle (E_\ell^B)^{4} \rangle$ & 10.63 & 10.69 & 10.58 & 0.16 & 0.16 & 0.16 \\
\bottomrule \bottomrule
\end{tabular}
}
\label{tab:charged-moments}

\end{table*}

\begin{table*}[!h]
\centering
\caption{Moment comparison for $B^0 \rightarrow X_u^- \ell^+ \nu_\ell$. DFN vs.~Bin-by-bin vs.~Optimal transport hybrids.}
\resizebox{\textwidth}{!}{
\begin{tabular}{lcccccc}
\toprule \toprule
& \multicolumn{3}{c}{Raw moments} & \multicolumn{3}{c}{Central moments} \\
\cmidrule(lr){2-4} \cmidrule(lr){5-7}
Observable & DFN & Bin-by-bin & Optimal transport & DFN & Bin-by-bin & Optimal transport \\
\midrule
$\langle (M_X^2)^{1} \rangle$ & 2.83 & 2.95 & 2.76 & 2.83 & 2.95 & 2.76 \\
$\langle (M_X^2)^{2} \rangle$ & 12.15 & 13.65 & 12.07 & 4.14 & 4.94 & 4.43 \\
$\langle (M_X^2)^{3} \rangle$ & 70.68 & 81.79 & 70.60 & 12.89 & 12.34 & 12.74 \\
$\langle (M_X^2)^{4} \rangle$ & 521.44 & 608.93 & 521.35 & 112.67 & 129.06 & 118.97 \\
\midrule
$\langle (q^2)^{1} \rangle$ & 6.66 & 6.33 & 6.70 & 6.66 & 6.33 & 6.70 \\
$\langle (q^2)^{2} \rangle$ & 67.08 & 62.78 & 68.36 & 22.76 & 22.70 & 23.46 \\
$\langle (q^2)^{3} \rangle$ & 829.05 & 788.59 & 861.73 & 79.53 & 103.66 & 89.26 \\
$\langle (q^2)^{4} \rangle$ & 11636.84 & 11556.03 & 12441.38 & 1503.76 & 1864.55 & 1712.31 \\
\midrule
$\langle (E_\ell^B)^{1} \rangle$ & 1.62 & 1.59 & 1.62 & 1.62 & 1.59 & 1.62 \\
$\langle (E_\ell^B)^{2} \rangle$ & 2.87 & 2.77 & 2.87 & 0.25 & 0.25 & 0.25 \\
$\langle (E_\ell^B)^{3} \rangle$ & 5.41 & 5.14 & 5.41 & -0.06 & -0.05 & -0.05 \\
$\langle (E_\ell^B)^{4} \rangle$ & 10.63 & 9.99 & 10.63 & 0.16 & 0.16 & 0.16 \\
\bottomrule \bottomrule
\end{tabular}
}
\label{tab:neutral-moments}

\end{table*}
\clearpage
\noindent \textbf{Data availability statement.} The simulation samples used in this analysis are publicly available on Zenodo~\cite{zenodo}.\\

\noindent \textbf{Code availability statement.} The code used in this analysis is publicly available on GitHub~\cite{github}.

\bibliographystyle{JHEP}
\bibliography{sn-bibliography}

\end{document}